\documentclass[12pt]{iopart}
\usepackage{graphicx}

\begin{document}

\title[]{A Modified Generalized Chaplygin Gas as the
Unified Dark Matter-Dark Energy Revisited}

\author{Xue-Mei Deng}

\address{Purple Mountain Observatory, Chinese Academy of Sciences,
Nanjing 210008, China} \ead{xmd@pmo.ac.cn}
\begin{abstract}
A modified generalized Chaplygin gas (MGCG) is considered as the
unified dark matter-dark energy revisited. The character of MGCG is
endued with the dual role, which behaves as matter at early times
and as an quiessence dark energy at late times. The equation of
state for MGCG is
$p=-\alpha\rho/(1+\alpha)-\vartheta(z)\rho^{-\alpha}/(1+\alpha) $,
where
$\vartheta(z)=-[\rho_{0c}(1+z)^{3}]^{(1+\alpha)}(1-\Omega_{0B})^{\alpha}\{\alpha\Omega_{0DM}+
\Omega_{0DE}[\omega_{DE}+\alpha(1+\omega_{DE})](1+z)^{3\omega_{DE}(1+\alpha)}\}$.
Some cosmological quantities, such as the densities of different
components of the universe $\Omega_{i}$ ($i$ respectively denotes
baryons, dark matter and dark energy) and the deceleration parameter
$q$, are obtained. The present deceleration parameter $q_{0}$ , the
transition redshift $z_{T}$ and the redshift $z_{eq}$, which
describes the epoch when the densities in dark matter and dark
energy are equal, are also calculated. To distinguish MGCG from
others, we then apply the Statefinder diagnostic. Later on, the
parameters ($\alpha$ and $\omega_{DE}$) of MGCG are constrained by
combination of the sound speed $c^{2}_{s}$, the age of the universe
$t_{0}$, the growth factor $m$ and the bias parameter $b$. It yields
$\alpha=-3.07^{+5.66}_{-4.98}\times10^{-2}$ and
$\omega_{DE}=-1.05^{+0.06}_{-0.11}$. Through the analysis of the
growth of density perturbations for MGCG, it is found that the
energy will transfer from dark matter to dark energy which reach
equal at $z_{eq}\sim 0.48$ and the density fluctuations start
deviating from the linear behavior at $z\sim 0.25$ caused by the
dominance of dark energy.
\end{abstract}

\pacs{98.80.-k.}

Keywords: Dark energy; Dark matter
\maketitle

\section{Introduction}

As standardized candles, type Ia supernovae suggest that the
universe is undergoing accelerating expansion\cite{SN Ia}. The same
evidence has been shown by cosmic microwave background
(CMB)\cite{CMB} and large scale structure (LSS)\cite{LSS}
observations. All these data indicate the dominant component of the
universe is relatively smooth and has a large negative pressure. All
kinds of alternative models stand up for explaining this exotic
phenomena within relativistic theory of gravity, which include the
energy of the quantum vacuum (such as cosmological constant
$\Lambda$\cite{L1,L2}), the existence of another new scalar field
(quintessence\cite{Q} which is possibly related to the inflaton,
k-essence\cite{k}, tachyon\cite{t}, phantom\cite{p}, and
quintom\cite{q}), and the influence of unseen additional spatial
dimensions predicted by string theory\cite{string}. Instead of the
presence of dark energy, the most interesting idea for solving this
dark riddle is depend on a new aspect of gravity which is not
accounted for general relativity\cite{gravity}. Unfortunately, the
current astronomical observations data still can not determine
completely the nature of dark energy or decide its existence since
this problem has puzzled us for a long time.

As one of plausible dynamical models for dark energy, Chaplygin gas
(CG) and the generalized Chaplygin gas (GCG)\cite{GCG}, has created
a lot of interest in recent times\cite{others}. CG or GCG behaves
like pressureless dust at early times and like a cosmological
constant during very late times. The fact that the properties of CG
and GCG interpolate between those of cold dark matter (CDM) and a
$\Lambda $-term makes the models to provide a conceptual framework
for a unified model of dark matter and dark energy. Besides, it is
more worthy of note that the GCG model has been successfully
confronted with various phenomenological tests such as high
precision CMB data, supernovae data and gravitational
lensing\cite{prove}. On the other hand, since observations such as
SNe Ia\cite{SN Ia} are still not sufficient to establish evidence of
a dynamical equation of state for dark energy, we can naturally
consider a fluid consisted of dark matter and quiessence dark energy
with constant $\omega_{DE}$ which full of the whole universe. Thus,
we propose a modified generalized Chaplygin gas (MGCG) as the
unified dark matter and dark energy revisited on some previous
works\cite{CG,MGCG,NGCG}.

With future deeper and more intensive surveys of SNe Ia, such as
Supernovae acceleration probe (SNAP)\cite{SNAP}, it provides a
chance to distinguish different models of dark energy so that we
will consider the so-called ``geometrical " or Statefinder
diagnostic since it can probe the expansion dynamics of the universe
through higher derivatives of the scale factor. So, we must
constrain the parameters ($\alpha$ and $\omega_{DE}$) of the MGCG at
first. The plat of this paper is as follows. In Sec. 2, we give
corresponding fundamental cosmology equations about MGCG.
Subsequently, in Sec. 3, we consider Statefinder diagnostic of the
MGCG. We then constrain the parameters of the MGCG by means of the
sound speed $c^{2}_{s}$, the age of the universe $t_{0}$, the growth
factor $m$ and the bias parameter $b$ in Sec. 4. Finally, conclusion
and our results are outlined in Sec. 5.

\section{Fundamental cosmology equations about MGCG}

\subsection{Motivation}

An interesting unified model for two dark sectors is the Chaplygin
gas (CG) model\cite{CGmodel} introduced with $p=-A/\rho$. Although
this model has been very successful in explaining the SNe Ia data,
it shows that CG model does not pass the tests connected with
structure formation and observed strong oscillations of the matter
power spectrum\cite{GW}. This situation can be alleviated in the
generalized Chaplygin gas (GCG) proposed with $p=-A/\rho^{\alpha}$.
And the parameter $\alpha$ is rather severely constrained, i.e.,
$0\leq\alpha<0.2$ at the $95\%$ confidence level\cite{GCG}. Since
the difference between $\Lambda $CDM and GCG models is so tiny and
the equation of state of dark energy still cannot be determined
exactly, a new generalized Chaplygin gas (NGCG)\cite{NGCG} proposed
by considering that $A$ in GCG is a function of redshift $z$.
Following this idea, we develop GCG-like models from another view.
As a purely kinetic k-essence model with a constant potential, the
tachyon fields can be considered as CG model \cite{CG}. Meanwhile
the tachyon field can also act as a source of dark energy depending
upon the form of the tachyon potential. Thus, the authors in Ref.
\cite{MGCG} introduced an extended tachyon field (ETF) and then
provided a modified Chaplygin gas (MCG)\cite{MGCG}, in which the
equation of state (EOS) has a more generalized form
\begin{equation}
\label{MCG}
p=-\frac{\alpha}{1+\alpha}\rho-\frac{1}{1+\alpha}\frac{A}{\rho^{\alpha}},
\end{equation}
where $A$ and $\alpha$ are constants. As a kind of new attempt just
like the extension from GCG to NGCG, we present a modified
generalized Chaplygin gas (MGCG) through replacing $A$ in Eq.
(\ref{MCG}) with a function $\vartheta(z)$ of redshift $z$. The
analysis and discussion of the matter power spectrum for MGCG will
be investigated in our future work by comparing the model with
observations.

\subsection{Deduction of some cosmological quantities and discussion}

Within the framework of Friedmann-Robertson-Walker (FRW) cosmology,
we consider an exotic background fluid called modified generalized
Chaplygin gas (MGCG) whose equation of state is as follows based on
Ref.\cite{MGCG}
\begin{equation}
\label{EOS}
p=-\frac{\alpha}{1+\alpha}\rho-\frac{1}{1+\alpha}\frac{\vartheta(z)}{\rho^{\alpha}},
\end{equation}
where $\alpha$ is a constant and $\vartheta(z)$ is a function of
redshift $z$. Since the MGCG is another new unified model with two
dark sectors (namely, at early times the energy density behaves as
dark matter: $\rho\propto(1+z)^{3}$; while at late times it behaves
like a quiessence dark energy:
$\rho\propto(1+z)^{3(1+\omega_{DE})}$), it could be supposed the
energy density of the MGCG as follows
\begin{equation}
\label{density}
\rho=[\kappa(1+z)^{3(1+\alpha)}+\lambda(1+z)^{3(1+\omega_{DE})(1+\alpha)}]^{\frac{1}{1+\alpha}},
\end{equation}
where the parameter of state for dark energy $\omega_{DE}$ is a
constant and should be taken as any value in the range $(-1.46,
-0.78)$\cite{w}. In Sec. 4, we will further constrain the
$\omega_{DE}$ in our model. Then the pressure of MGCG yields
\begin{equation}
\label{p}
p=-\frac{\alpha[\kappa(1+z)^{3(1+\alpha)}+\lambda(1+z)^{3(1+\omega_{DE})(1+\alpha)}]+\vartheta(z)}
{(1+\alpha)[\kappa(1+z)^{3(1+\alpha)}+\lambda(1+z)^{3(1+\omega_{DE})(1+\alpha)}]^{\alpha/(1+\alpha)}}.
\end{equation}

On the other hand, the whole pressure $p_{total}$ and energy density
$\rho_{total}$ satisfy the conservation equation
\begin{equation}
\label{total density}
\dot{\rho}_{total}+3H(\rho_{total}+p_{total})=0.
\end{equation}
According to conservation of baryons, the continuity equation is
separated into two parts:
\begin{equation}
\label{CEB} \dot{\rho}_{B}+3H(\rho_{B}+p_{B})=0,
\end{equation}
\begin{equation}
\label{CE} \dot{\rho}+3H(\rho+p)=0,
\end{equation}
where subscript ``B" denotes baryons. Using the Eqs.
(\ref{density}), (\ref{p}) and (\ref{CE}), we obtain
\begin{eqnarray}
\label{extended function}
\vartheta(z)&&=-\bigg\{[\omega_{DE}+\alpha(1+\omega_{DE})]\lambda(1+z)^{3(1+\omega_{DE})(1+\alpha)}\nonumber\\
&&+\alpha\kappa(1+z)^{3(1+\alpha)}\bigg\}.
\end{eqnarray}

Furthermore, the pressure of dark energy $p_{DE}$, the energy
density of dark energy $\rho_{DE}$ and the one of dark matter
$\rho_{DM}$ can be respectively written as
\begin{eqnarray}
\label{PDE}
p_{DE}&&=p\nonumber\\
&&=\frac{\omega_{DE}\lambda(1+z)^{3(1+\omega_{DE})(1+\alpha)}}{[\kappa(1+z)^{3(1+\alpha)}+\lambda(1+z)^{3(1
+\omega_{DE})(1+\alpha)}]^{\alpha/(1+\alpha)}},
\end{eqnarray}
\begin{eqnarray}
\label{rhoDE}
\rho_{DE}&&=\frac{p_{DE}}{\omega_{DE}}\nonumber\\
&&=\frac{\lambda(1+z)^{3(1+\omega_{DE})(1+\alpha)}}{[\kappa(1+z)^{3(1+\alpha)}+\lambda(1+z)^{3(1+\omega_{DE})(1+\alpha)}]^{\alpha/(1+\alpha)}},
\end{eqnarray}
\begin{eqnarray}
\label{rhoDM}
\rho_{DM}&&=\rho-\rho_{DE}\nonumber\\
&&=\frac{\kappa(1+z)^{3(1+\alpha)}}{[\kappa(1+z)^{3(1+\alpha)}+\lambda(1+z)^{3(1+\omega_{DE})(1+\alpha)}]^{\alpha/(1+\alpha)}}.
\end{eqnarray}
From Eqs. (\ref{density}), (\ref{rhoDE}) and (\ref{rhoDM}), it gives
that
\begin{equation}
\kappa+\lambda=\rho_{0}^{1+\alpha}=(\rho_{0DM}+\rho_{0DE})^{1+\alpha}=\rho_{0C}^{1+\alpha}[1-\Omega_{0B}]^{1+\alpha},
\end{equation}
\begin{equation}
\rho_{0DM}=\frac{\kappa}{[\kappa+\lambda]^{\alpha/(1+\alpha)}},
\end{equation}
\begin{equation}
\rho_{0DE}=\frac{\lambda}{[\kappa+\lambda]^{\alpha/(1+\alpha)}},
\end{equation}
where $\rho_{0C}$, $\rho_{0}$, $\rho_{0DM}$ and $\rho_{0DE}$ are the
present values of $\rho_{total}$, $\rho$, $\rho_{DM}$ and
$\rho_{DE}$, respectively and the label ``0" denotes today's
evaluated quantities. Parameters $\kappa$ and $\lambda$ can then be
written as
\begin{equation}
\label{constant}
\kappa=\Omega_{0DM}\rho_{0C}^{1+\alpha}[1-\Omega_{0B}]^{\alpha},
~~~~\lambda=\Omega_{0DE}\rho_{0C}^{1+\alpha}[1-\Omega_{0B}]^{\alpha}.
\end{equation}
Substituting Eq. (\ref{constant}) to Eq. (\ref{extended function}),
we have
\begin{eqnarray}
\label{theta} \vartheta(z)&&=
-\rho_{0C}^{1+\alpha}[1-\Omega_{0B}]^{\alpha}(1+z)^{3(1+\alpha)}\bigg\{\alpha\Omega_{0DM}\nonumber\\
&&+\Omega_{0DE}[\omega_{DE}+\alpha(1+\omega_{DE})](1+z)^{3\omega_{DE}(1+\alpha)}\bigg\}.
\end{eqnarray}

Then, the equation of state for the MGCG model \textbf{reads as}
\begin{equation}
\label{eos}
\chi(z)\equiv\frac{p}{\rho}=\frac{\omega_{DE}\Omega_{0DE}(1+z)^{3(1+\omega_{DE})(1+\alpha)}}{\Omega_{0DM}(1+z)^{3(1+\alpha)}
+\Omega_{0DE}(1+z)^{3(1+\omega_{DE})(1+\alpha)}}.
\end{equation}

\begin{figure}[htbp]
\includegraphics[height=12cm,angle=270]{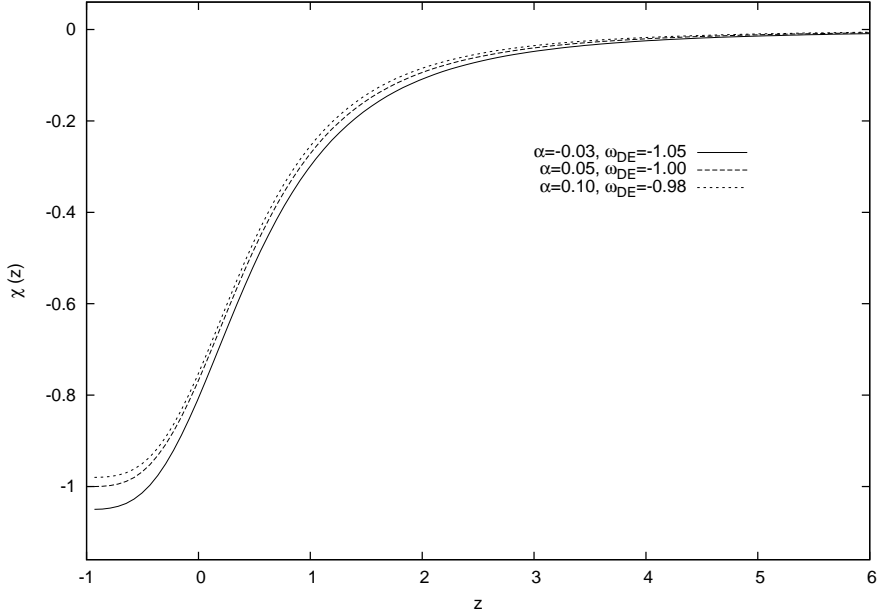}
 \caption{\label{fig2}The equation of state for the MGCG $\chi(z)$
as the function of redshift. Where $\Omega_{0B}=0.0449$,
$\Omega_{0DM}=0.222$ and $\Omega_{0DE}=0.734$.}
\end{figure}

The values of $\Omega_{0DM}$ and $\Omega_{0DE}$ depend on the model
used in the data precessing. For example, both the Wilknson
Microware Anisotropy Probe (WMAP)\cite{WMAP} and two-degree field
Galaxy Redshift Survey (2dFGRS)\cite{2dF} used $\Lambda$CDM model of
the universe. Besides, cosmological parameters are also constrained
with other models. In this papar, the current density parameters
used in the plots are $\Omega_{0B}=0.0449\pm0.0028$,
$\Omega_{0DM}=0.222\pm0.026$ and $\Omega_{0DE}=0.734\pm0.029$ based
on WMAP7 data\cite{WMAP}. Later, we will give the constraints of
$\alpha$ and $\omega_{DE}$ in section 4. From Eq.(\ref{eos}), we can
plot the function $\chi(z)$ as the function of redshift z ( see Fig.
\ref{fig2}). $\chi(z)$ is always negative from past to future. And
we can see the effect of the parameter $\alpha$ and $\omega_{DE}$ on
the $\chi(z)$. The evolution of $\chi(z)$ is mostly flat in the high
redshift $z>1$ and is very steep as redshift $z$ becomes low.

\begin{figure}[htbp]\center
\includegraphics[height=12cm,angle=270]{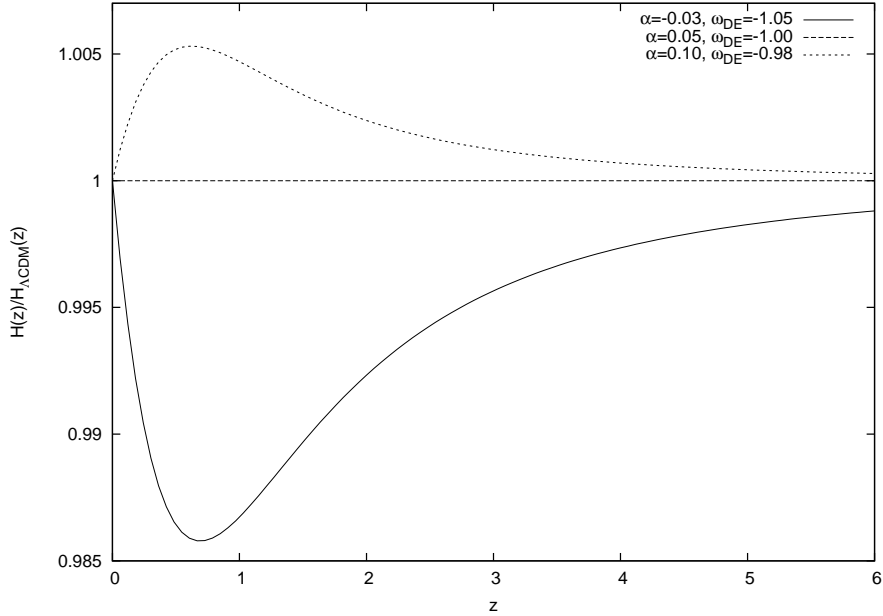}
 \caption{\label{fig3}The Hubble parameter in units of $H_{\Lambda
CDM}$ for the MGCG as the function of redshift. The priors
$\Omega_{0DM}=0.222$, $\Omega_{0DE}=0.734$ and $\Omega_{0B}=0.0449$
have been used.}
\end{figure}

When we consider a spatially flat FRW universe with the exotic
background MGCG fluid and the baryon component, the Friedmann
equation can be written as:
\begin{equation}
H^{2}=\frac{8\pi G}{3}\rho_{total}.
\end{equation}
Then, we obtain
\begin{eqnarray}
E^{2}(z)&&\equiv\frac{H^{2}}{H^{2}_{0}}\nonumber\\
&&=(1-\Omega_{0B})^{\frac{\alpha}{1+\alpha}}[\Omega_{0DM}(1+z)^{3(1+\alpha)}
+\Omega_{0DE}(1+z)^{3(1+\omega_{DE})(1+\alpha)}]^{\frac{1}{1+\alpha}}\nonumber\\
&&+\Omega_{0B}(1+z)^{3}.
\end{eqnarray}
The evolution of the Hubble parameter in units of
$H_{\Lambda\mathrm{ CDM}}$ is plotted in Fig. \ref{fig3}. During the
cosmological evolution, the behavior of $H$ is similar to
$H_{\Lambda\mathrm{CDM}}$ at very early time, and when $z\rightarrow
0$, $H$ is approaching $\Lambda$CDM.

\begin{figure}[htbp]\center
\includegraphics[height=12cm,angle=270]{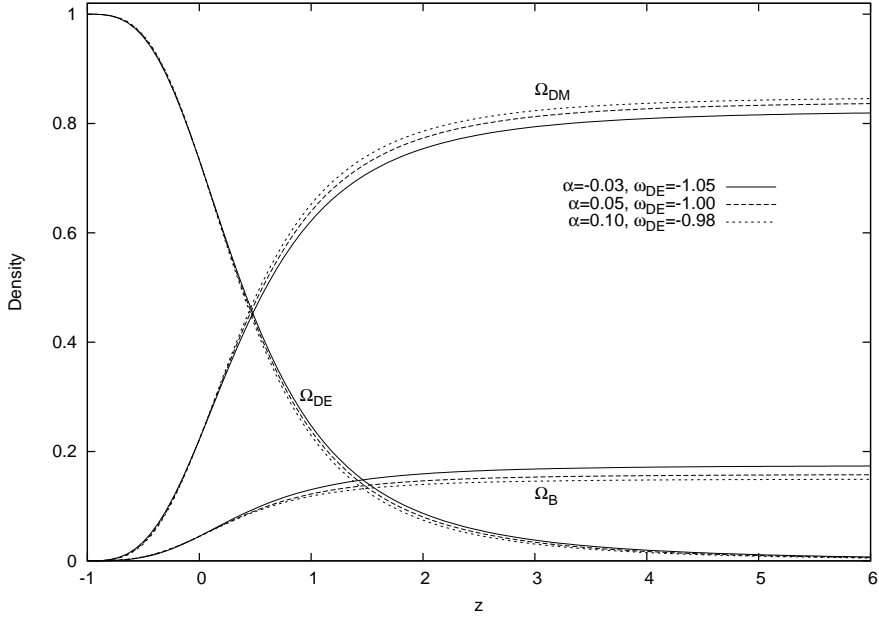}
 \caption{\label{fig4}The densities of $\Omega_{DE}$, $\Omega_{DM}$
and $\Omega_{B}$ as the functions of redshift. }
\end{figure}

By using the above equations, the densities of different components
of the universe $\Omega_{DM}$, $\Omega_{DE}$ and $\Omega_{B}$ can be
respectively derived as
\begin{equation}
\Omega_{DM}=\frac{(1-\Omega_{0B})^{\frac{\alpha}{1+\alpha}}\Omega_{0DM}(1+z)^{3}}{E^{2}(z)[\Omega_{0DM}
+\Omega_{0DE}(1+z)^{3\omega_{DE}(1+\alpha)}]^{\alpha/(1+\alpha)}},
\end{equation}
\begin{equation}
\Omega_{DE}=\frac{(1-\Omega_{0B})^{\frac{\alpha}{1+\alpha}}\Omega_{0DE}(1+z)^{3(1+\omega_{DE})}}{E^{2}(z)[\Omega_{0DM}(1+z)^{-3\omega_{DE}(1+\alpha)}
+\Omega_{0DE}]^{\alpha/(1+\alpha)}},
\end{equation}
\begin{equation}
\Omega_{B}=\frac{\Omega_{0B}(1+z)^{3}}{E^{2}(z)}.
\end{equation}
The relationships between $\Omega _{i}$ ($i$ respectively denotes
baryons, dark matter and dark energy) and redshift $z$ are shown in
Fig. \ref{fig4} with the prior of $\Omega _{0DM}=0.222$, $\Omega
_{0DE}=0.734 $ and $\Omega _{0B}=0.0449$. Based on Fig. \ref{fig4},
the evolutions of baryons, dark matter and dark energy under MGCG
model are consistent with what are recognized\cite{w,evolution}. An
very important redshift described the epoch when the densities in
dark matter and dark energy are equal is
$(1+z_{eq})^{-3\omega_{DE}(1+\alpha)}=\Omega_{0DE}/\Omega_{0DM}$.
And \textbf{$z_{eq}\simeq0.48$} for
$\alpha=-0.03,\omega_{DE}=-1.05$, \textbf{$z_{eq}\simeq0.46$} for
$\alpha=0.05,\omega_{DE}=-1.00$, \textbf{$z_{eq}\simeq0.45$} for
$\alpha=0.10,\omega_{DE}=-0.98$ respectively. From Fig. \ref{fig4}
we also know that the dominance of the dark energy leads to the
acceleration expansion of our universe. The increasing density of
dark energy is the reason why expansion of our universe transited
from deceleration to acceleration. This transition will be described
by our model in more detail with the deceleration parameter $q$.

\begin{figure}[htbp]\center
\includegraphics[height=12cm,angle=270]{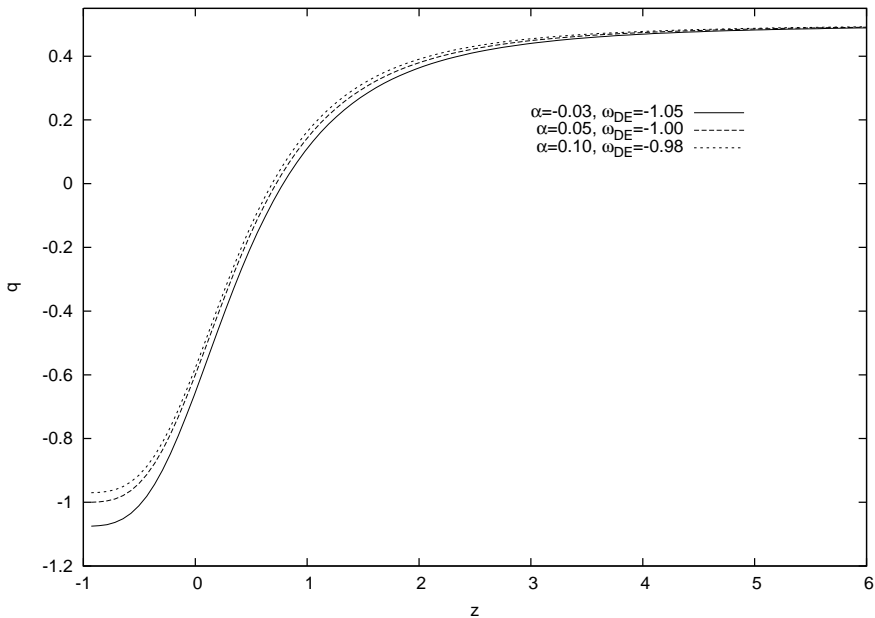}
 \caption{\label{fig5}The deceleration parameter $q$ as the function
of redshift. The priors $\Omega_{0DM}=0.222$, $\Omega_{0DE}=0.734$
and $\Omega_{0B}=0.0449$ have been used.}
\end{figure}

The deceleration parameter is
\begin{eqnarray}
q&&\equiv-\frac{\ddot{a}}{aH^{2}}\nonumber\\
&&=-1-\frac{\dot{H}}{H^{2}}\nonumber\\
&&=-1+\frac{3(1+z)^{3}}{2E^{2}(z)}(1-\Omega_{0B})^{\alpha/(1+\alpha)}\bigg\{[\Omega_{0DM}
+\Omega_{0DE}(1+z)^{3\omega_{DE}(1+\alpha)}]^{-\frac{\alpha}{1+\alpha}}\nonumber\\
&&\times[\Omega_{0DM}+(1+\omega_{DE})\Omega_{0DE}(1+z)^{3\omega_{DE}(1+\alpha)}]+\Omega_{0B}\bigg\}.
\end{eqnarray}
Thus we can give the following results: the present deceleration
parameter $q_0\approx -0.65$ according to $z=0$ and the transition
redshift $z_T\approx 0.70$ according to $q=0$ when taking
$\alpha=-0.03$, $\omega _{DE}=-1.05$, $q_0\approx -0.60$ and
$z_T\approx 0.65$ for $\alpha=0.05$, $\omega _{DE}=-1.00$, and
$q_0\approx -0.58$ and $z_T\approx 0.60$ for $\alpha=0.10$, $\omega
_{DE}=-0.98$, which are consistent with observations. The
deceleration parameter as the function of redshift is shown in Fig.
\ref{fig5}. According to Fig. \ref{fig5}, the expansion of the
universe is from slowing down $(q>0)$ in the past to speeding up
$(q<0)$ at present time and in the future. And when $q>0$ is at the
high redshift, $q$ changes very slow with redshift $z$; when $q<0$
is at the recent time, and $q$ changes faster and faster
with redshift $z$.

Now, we compare some cosmological characters between CG, GCG, NGCG,
MCG and MGCG (see Table \ref{compare}). In order to distinguish
these model, we use the same symbols as above. From Table
\ref{compare}, we can see the EOSs of these models so that they have
some different characters. First of all, CG and GCG can be treated
as a unified fluid whose behaves like pressureless dust at early
times and like a cosmological constant during very late times. For
NGCG, it is considered as $X$-CDM, where $X$ means quiessence, since
the authors in Ref. \cite{NGCG} extended $A$ of GCG model. And MCG
is the mixture of a barotropic perfect fluid and a generalized
Chaplygin gas. At the same time, considering $A$ of Eq. (\ref{MCG})
changed with redshift $z$, we proposed MGCG.

\begin{table}[t]
\caption{\label{compare} Compare CG, GCG, NGCG, MCG and MGCG}
\begin{tabular}{lllll}
\br
Models & EOS & A & Parameter(s) & $\rho_{DM}/\rho_{DE}$\\
\mr
CG&$p=-\frac{A}{\rho}$&Const.&$A$&$\sim(1+z)^{3}$\\
GCG&$p=-\frac{A}{\rho^{\alpha}}$&Const.&$A$ and $\alpha$&$\sim(1+z)^{3(1+\alpha)}$\\
NGCG&$p=-\frac{\tilde{A}(z)}{\rho^{\alpha}}$&Function of $z$&$\omega_{DE}$ and $\alpha$&$\sim(1+z)^{-3\omega_{DE}(1+\alpha)}$\\
MCG&$p=-\frac{\alpha}{1+\alpha}\rho-\frac{1}{1+\alpha}\frac{A}{\rho^{\alpha}}$&Const.&$A$ and $\alpha$&$\sim(1+z)^{3(1+\alpha)}$\\
MGCG&$p=-\frac{\alpha}{1+\alpha}\rho-\frac{1}{1+\alpha}\frac{\vartheta(z)}{\rho^{\alpha}}$&Function
of $z$&$\omega_{DE}$
 and $\alpha$&$\sim(1+z)^{-3\omega_{DE}(1+\alpha)}$\\
\br
\end{tabular}
\end{table}

Among them, GCG is a extended model of CG. When $\alpha=1$, GCG
model becomes CG. And NGCG is a extended model of GCG. When
$\tilde{A}(z)=const.$, NGCG reduces to GCG. MGCG is a extended
version of MCG. When $\vartheta(z)=const.$, MGCG yields MCG.
Meanwhile, both GCG and MCG can be conceived as ETF, in which GCG
corresponds to the simplest model driven by a constant
potential\cite{MGCG}. It is worth of noted that CG, GCG, NGCG, MCG
and MGCG are all interact models. Namely, there are energy exchange
between dark matter and dark energy. Basically, the behaviour of the
energy exchange can be presented as $\rho_{DM}/\rho_{DE}$. We can
see the transfer direction of the energy flow for CG is from dark
matter to dark energy. However, the transfer direction of the energy
flow for the others only depend on the parameter $\alpha$ if
$\alpha\neq0$. When $\alpha>0$, the transfer direction of the energy
flow is from dark matter to dark energy. When $\alpha<0$, the
direction of transfer is inverse. Thus, constraints of parameters
such as $\alpha$ and $\omega_{DE}$ from observational data and
theoretical analysis will play an important role in understanding
the physical nature of these models. For GCG, authors in
Ref.\cite{a} obtain $\alpha<10^{-5}$ or $\alpha \geq 3$ (the sound
speed is larger than the speed of light) by using gauge-invariant
analysis of perturbation. For NGCG, constraints of parameters are
$1+\alpha=1.06^{+0.20}_{-0.16}$ and
$\omega_{DE}=-0.98^{+0.15}_{-0.20}$ by observational constraints
from SNe Ia, CMB and LSS data\cite{NGCG}. In section 4, we will
constrain $\alpha$ and $\omega_{DE}$ in our model by the maximum
likelihood estimation.

\section{Statefinder diagnostic}

Dark energy models such as the cosmological constant\cite{L1,L2},
quintessence\cite{Q}, K-essence\cite{k}, Chaplygin gas\cite{GCG} and
quintom\cite{q} etc. have properties which can be model-dependent.
In order to distinguish the very distinct and competing cosmological
scenarios involving dark energy, a sensitive and robust diagnostic
of dark energy is needed. So a new diagnostic of dark energy called
``Statefinder" diagnostic has been constructed by Sahni $et$
$al.$\cite{statefinder} who were using both the second and the third
derivatives of the scale factor $a$.

The statefinder pair $\{r, s\}$, in addition to the oldest and most
well-known geometric variables $H$ and $q$, defines two new
cosmological parameters
\begin{equation}
\label{r} r\equiv\frac{\stackrel{...}{a}}{aH^{3}},
\end{equation}
\begin{equation}
\label{s} s\equiv\frac{r-1}{3(q-1/2)}.
\end{equation}
An important property of the Statefinder is that spatially flat
$\Lambda$CDM corresponds to the fixed point
\begin{equation}
    \{r,s\}|_{\Lambda\mathrm{CDM}}=\{1,0\}.
\end{equation}
Clearly an important requirement of any diagnostic is that it
permits us to tell difference between a given dark energy model and
the simplest of all models - the cosmological constant just as
demonstrated in\cite{statefinder}-\cite{other statefinder}. By using
the $r(s)$ evolution diagram, the discrimination between a given
dark energy model and the $\Lambda$CDM scenario can be clearly
identified. And it is more worth of noted that the Statefinder
diagnostic combined with future SNAP\cite{SNAP} observations may
possibly be used to distinguish between different dark energy
models.

\begin{figure}[htbp]\center
\includegraphics[height=12cm,angle=270]{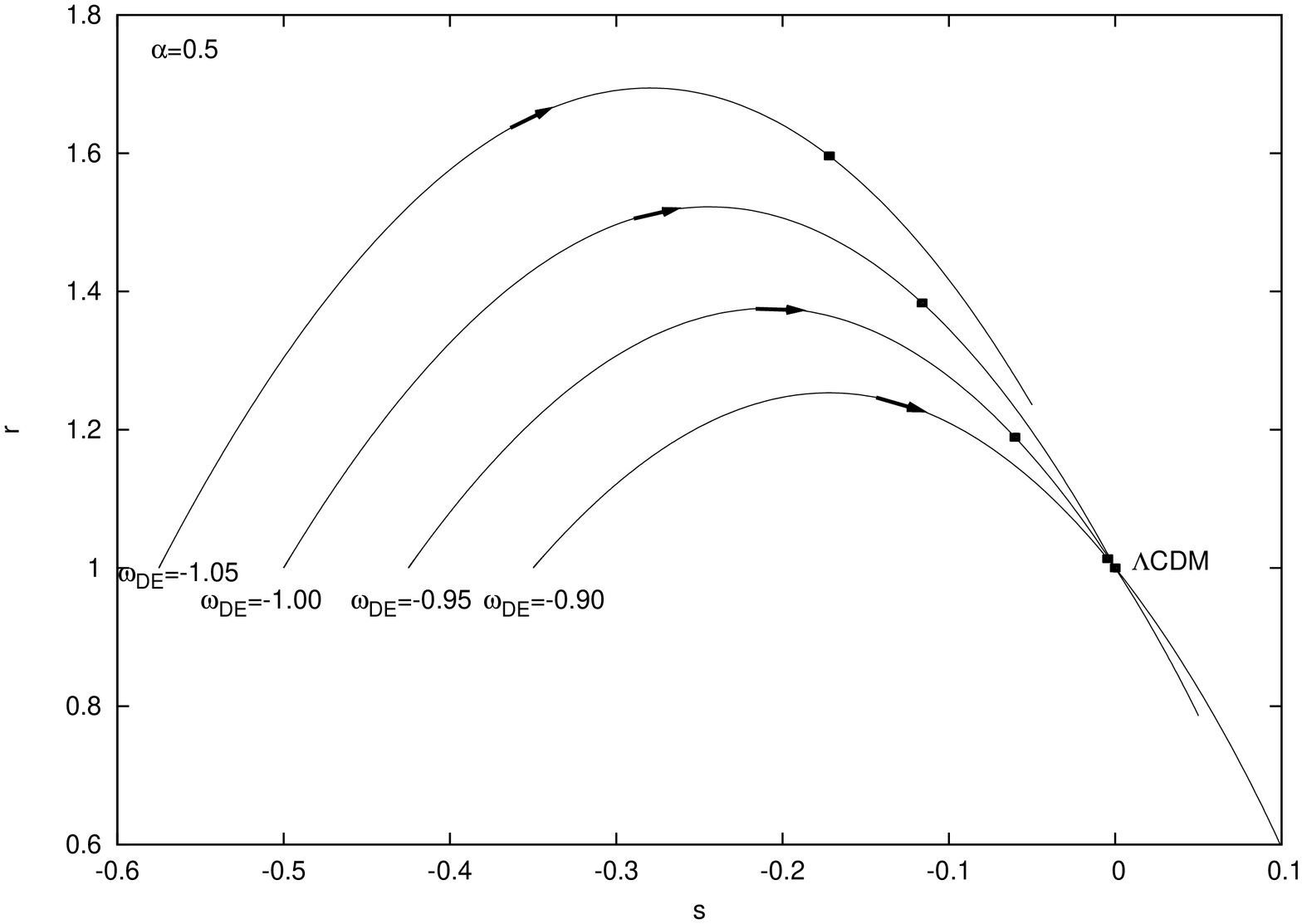}
 \caption{\label{fig6}The Statefinder $r(s)$ evolution diagram when
we fix $\alpha$ and vary $\omega_{DE}$. Dots locate today's values
and arrows denote the direction of evolution for the MGCG.}
\end{figure}

\begin{figure}[htbp]\center
\includegraphics[height=12cm,angle=270]{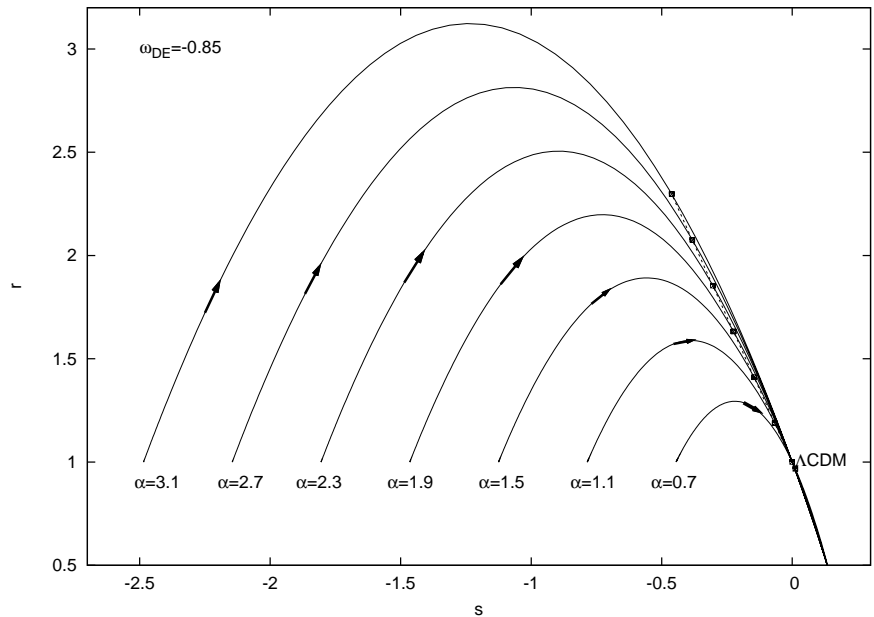}
 \caption{\label{fig7}The Statefinder $r(s)$ evolution diagram when
we fix $\omega_{DE}$ and vary $\alpha$. Dots locate today's values
and arrows denote the direction of evolution for the MGCG.}
\end{figure}

Based on the Eqs. (\ref{r}) and (\ref{s}), we obtain the Statefinder
parameters for the MGCG as follows
\begin{equation}
\label{R}
r=1+3\frac{\dot{H}}{H^{2}}+\frac{\ddot{H}}{H^{3}}\equiv1+3\Upsilon(z)+\Theta(z),
\end{equation}
\begin{equation}
\label{S}
s=-\frac{6\dot{H}/H^{2}+2\ddot{H}/H^{3}}{9+6\dot{H}/H^{2}}\equiv-\frac{6\Upsilon(z)+2\Theta(z)}{9+6\Upsilon(z)},
\end{equation}
where
\begin{eqnarray}
\Upsilon(z)&&=\frac{\dot{H}}{H^{2}}\nonumber\\
&&=-\frac{3}{2}-\frac{3}{2}\omega_{DE}\Omega_{0DE}(1+z)^{3\omega_{DE}(1+\alpha)}(1-\Omega_{0B})^{\alpha/(1+\alpha)}
\bigg\{(1-\Omega_{0B})^{\alpha/(1+\alpha)}\nonumber\\
&&\times\bigg[\Omega_{0DM}+\Omega_{0DE}(1+z)^{3\omega_{DE}(1+\alpha)}\bigg]+\Omega_{0B}\bigg[\Omega_{0DM}\nonumber\\
&&+\Omega_{0DE}(1+z)^{3\omega_{DE}(1+\alpha)}\bigg]^{\alpha/(1+\alpha)}\bigg\}^{-1},
\end{eqnarray}
and
\begin{eqnarray}
\Theta(z)&&=\frac{\ddot{H}}{H^{3}}\nonumber\\
&&=\frac{9}{2}+\frac{9}{2}(1-\Omega_{0B})^{\alpha/(1+\alpha)}\bigg\{2\omega_{DE}\Omega_{0DE}(1+z)^{3\omega_{DE}(1+\alpha)}\nonumber\\
&&+(1+\alpha)\omega^{2}_{DE}\Omega_{0DE}(1+z)^{3\omega_{DE}(1+\alpha)}\nonumber\\
&&-\alpha\omega^{2}_{DE}\Omega^{2}_{0DE}(1+z)^{6\omega_{DE}(1+\alpha)}\bigg[\Omega_{0DM}\nonumber\\
&&+\Omega_{0DE}(1+z)^{3\omega_{DE}(1+\alpha)}\bigg]^{-1}\bigg\}\bigg\{(1-\Omega_{0B})^{\alpha/(1+\alpha)}\nonumber\\
&&\times\bigg[\Omega_{0DM}+\Omega_{0DE}(1+z)^{3\omega_{DE}(1+\alpha)}\bigg]\nonumber\\
&&+\Omega_{0B}\bigg[\Omega_{0DM}
+\Omega_{0DE}(1+z)^{3\omega_{DE}(1+\alpha)}\bigg]^{\alpha/(1+\alpha)}\bigg\}^{-1}.
\end{eqnarray}
Then, the evolution trajectories of our model in the plane of the
Statefinder parameters can also be plotted (see Fig. \ref{fig6} and
Fig. \ref{fig7}).

Since different dark energy models have distinct evolution
trajectories in the plane of the Statefinder parameters, we can
distinguish between various dark energy models and the MGCG. Just as
the literatures\cite{statefinder,other statefinder} show: the
$\Lambda$CDM scenario corresponds to the fixed point $\{r =1,s = 0
\}$; the standard cold dark matter (SCDM) scenario corresponds to
the point $\{ r = 1,s = 1\}$; the Statefinder parameters plane of
quiessence are some vertical segments where $r$ decreases
monotonically from $1$ to $1+ \frac{9}{2}\omega_{DE}(1 +
\omega_{DE})$ and $s$ remains constant at $1 + \omega_{DE}$; the
quintessence tracker models have typical trajectories similar to
arcs of an upward parabola lying in the regions $\{ r < 1,s > 0\}$.

Now, we can see the Statefinder parameters plane of the MGCG from
Fig. \ref{fig6} and Fig. \ref{fig7}. In Fig. \ref{fig6}, we fix
$\alpha=0.5$ and vary $\omega_{DE}$ as $-1.05$, $-1.00$, $-0.95$ and
$-0.90$ respectively. Where $\omega_{DE}=-1.00$ exhibits a complete
downward parabola, while the others correspond to some broken
parabolas. And from this diagram, we can see the Statefinder
trajectory begins with $r=1$. Besides, the cases of $\omega_{DE}<-1$
never arrive at the fixed point $\{1,0\}$ and when $\omega_{DE}>-1$
the trajectories must pass through this point. Finally, we can see
that the less $\omega_{DE}$, the smaller today's value $s_{0}$ and
the larger today's value $r_{0}$. However, in Fig. \ref{fig7}, we
fix $\omega_{DE}=-0.85$ and vary $\alpha$ as $3.1$, $2.7$, $2.3$,
$1.9$, $1.51$, $1.1$ and $0.7$ respectively. It is interesting to
see that the trajectories can pass through the fixed point
$\{1,0\}$. And when fixed $\omega_{DE}$, the less $\alpha$, the
smaller today's value $s_{0}$ and the larger today's value $r_{0}$.
We notice the present Statefinder points as well as the fixed point
$\{1,0\}$ locate on a straight line which means the relationship
between $r_{0}$ and $s_{0}$ is linear when we fix $\omega_{DE}$.
That is because the following relationships
\begin{eqnarray}
r_{0}&&=1+\frac{9}{2}\omega_{DE}\Omega_{0DE}\bigg[1+(1+\alpha)\omega_{DE}-\frac{\alpha\omega_{DE}
\Omega_{0DE}}{1-\Omega_{0B}}\bigg],\nonumber\\
s_{0}&&=1+(1+\alpha)\omega_{DE}-\frac{\alpha\omega_{DE}\Omega_{0DE}}{1-\Omega_{0B}},
\end{eqnarray}
so that
\begin{equation}
r_{0}=1+\frac{9}{2}\omega_{DE}\Omega_{0DE}s_{0}.
\end{equation}

Through above, the separation between distinct families of dark
energy models is very remarkable when we analyze evolutionary
trajectories using the Statefinder pair $\{r,s\}$. Forthcoming
space-based missions such as SNAP are expected to greatly increase
and improve the current Type Ia supernovae inventory, and maybe can
identify which kind of dark energy model.

\section{Constraints on MGCG}

\subsection{The sound speed}
\begin{figure}[htbp]\center
\includegraphics[height=12cm,angle=270]{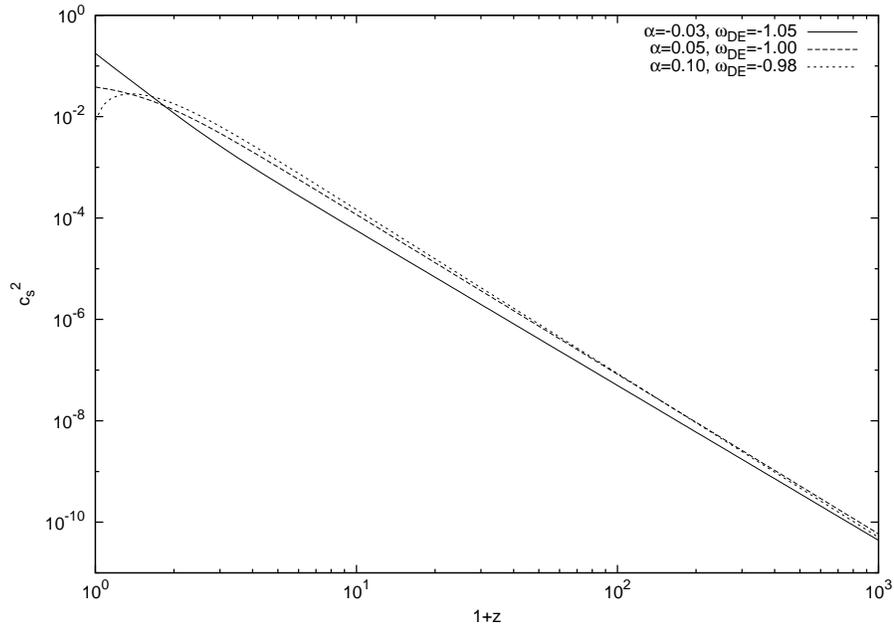}
 \caption{\label{fig8}The evolution of the sound speed squared of
the MGCG as a function of redshift for $\alpha=-0.03$, $0.05$ and
$0.10$.}
\end{figure}

Since our model can be considered as a new scenario for unified dark
matter and dark energy, we have to study its density perturbations
and the structure formation, we also have to investigate its
adiabatic sound speed. So, in this section, we consider some
effective parameters for the MGCG such as $\alpha$ and $\omega_{DE}$
by the maximum likelihood estimation. If the MGCG is considered as a
perfect fluid satisfying equation (\ref{EOS}), then the MGCG
component will cluster gravitationally with the adiabatic sound
speed given by
\begin{figure}[htbp]\center
\includegraphics[height=12cm,angle=270]{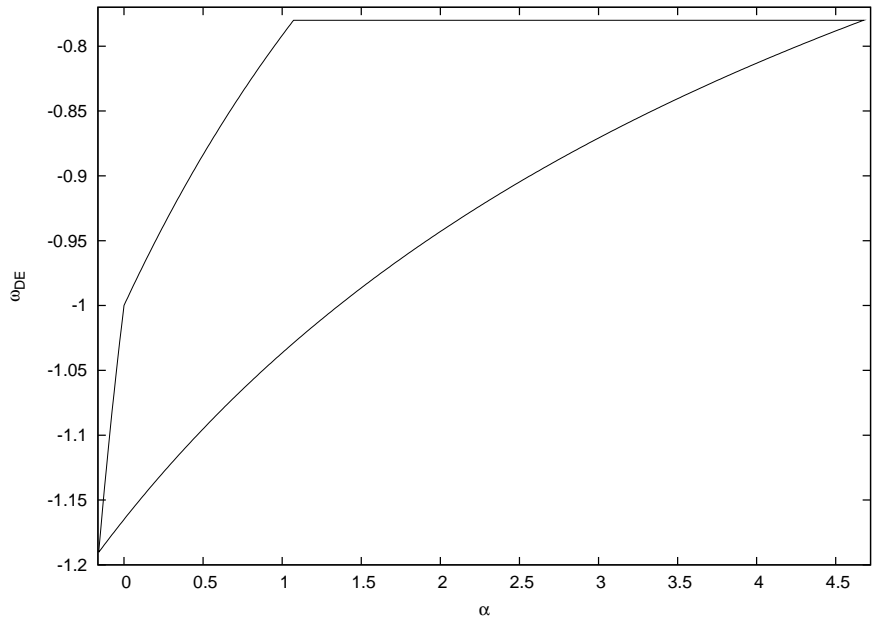}
 \caption{\label{fig9}The region inside the curve is the domain of
parameters $\alpha$ and $\omega_{DE}$ which is constraints from the
sound speed.}
\end{figure}

\begin{eqnarray}
\label{sound speed}
c^{2}_{s}&&\equiv\frac{\partial p}{\partial\rho}\nonumber\\
&&=-\frac{\alpha}{1+\alpha}-\frac{\alpha}{1+\alpha}\frac{\Omega_{0DE}[\omega_{DE}+\alpha(1+\omega_{DE})](1+z)^{3\omega_{DE}(1+\alpha)}
+\alpha\Omega_{0DM}}
{\Omega_{0DE}(1+z)^{3\omega_{DE}(1+\alpha)}+\Omega_{0DM}}\nonumber\\
&&+\frac{(1+\omega_{DE})\Omega_{0DE}[\omega_{DE}+\alpha(1+\omega_{DE})](1+z)^{3\omega_{DE}(1+\alpha)}+\alpha\Omega_{0DM}}
{(1+\omega_{DE})\Omega_{0DE}(1+z)^{3\omega_{DE}(1+\alpha)}+\Omega_{0DM}}.
\end{eqnarray}

From the above equation, we obtain $\alpha\neq-1$. However, when
$\alpha<-1$, we can derive
$c^{2}_{s}|_{z\rightarrow\infty}\simeq\omega_{DE}<0$, and when
$c^{2}_{s}$ is negative, the MGCG fluid is unstable. When
$\alpha>-1$, we have $c^{2}_{s}|_{z\rightarrow\infty}\simeq0$. Thus,
the parameter $\alpha$ must be greater than $-1$ because the limit
of the sound speed was very small at early time and can not be
larger than $1$ even though in the future if our model can be as the
unified dark matter-dark energy. From Fig. \ref{fig8}, we can see
the slope of the curves for $c^{2}_{s}$ depends on $\alpha$ and
$\omega_{DE}$. Besides, we clearly see that deep in the matter era
the behavior of the MGCG closely resembles that of CDM.

To constrain $\alpha$ and $\omega_{DE}$ further, combined with the
data of SNe Ia, CMB and 2dFGRS, it imposes a range on $\omega_{DE}$:
$-1.46<\omega_{DE}<-0.78$\cite{w}. Through the limit of the sound
speed $0<c^{2}_{s}<1$, when we use the Eq. (\ref{sound speed}), we
can find the domain of $\alpha$ - $\omega_{DE}$ plane (see Fig.
\ref{fig9}). Then, we derive $-0.3<\alpha<4.7$ and
$-1.19<\omega_{DE}<-0.78$. From Fig. \ref{fig9}, we can see the
boundary of domain is constraints from the sound speed.

\subsection{The age of the universe}
\begin{figure}[htbp]\center
\includegraphics[height=9cm,angle=0]{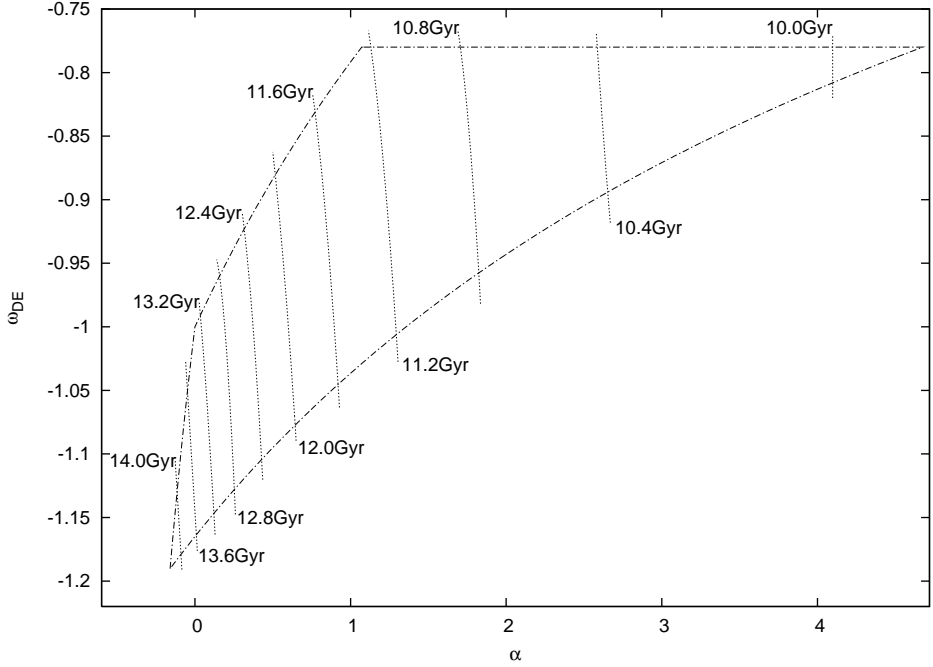}
 \caption{\label{fig10}The dashed-dot curve is the domain of
parameters $\alpha$ and $\omega_{DE}$ which is constraints from the
sound speed. The dot lines is the isochrones for parameter $t_{0}$
in the $\alpha$-$\omega_{DE}$ plane.}
\end{figure}

By using a distance-independent method, Jimenez et al.\cite{age1}
determined the age of globular clusters in the Milky Way as
$t_{0}=13.5\pm2$Gyr. By using the white dwarfs cooling sequence
method, Richer et al.\cite{age2} and Hansen et al.\cite{age3}
constrained the age of the globular cluster M4 to be
$t_{0}=12.7\pm0.7$Gyr. Then, the age of the universe needs to
satisfy the lower bound: $t_{0}>11-12$Gyr. Assuming $\Lambda$CDM,
WMAP7 data give the age of the universe
$t_{0}=13.73\pm0.13$Gyr\cite{WMAP}. Here we adopt model-independent
astronomical observations of globular clusters as a criterion.
Making use of the definition for the Hubble parameter $H=\dot{a}/a$
and the relationship between the scale factor and the redshift
$1+z=1/a$, then we have
\begin{equation}
dt=\frac{da}{aH}=-\frac{1}{(1+z)H}dz.
\end{equation}
Integrated the above equation, we obtain
\begin{eqnarray}
\label{time}
t_{0}&&=\int^{t_{0}}_{0}dt=\int^{\infty}_{0}\frac{dz}{H(1+z)}\nonumber\\
&&=\frac{1}{H_{0}}\int^{\infty}_{0}\frac{dz}{(1+z)^{5/2}\sqrt{(1-\Omega_{0B})^{\frac{\alpha}{1+\alpha}}[\Omega_{0DM}
+\Omega_{0DE}(1+z)^{3\omega_{DE}(1+\alpha)}]^{\frac{1}{1+\alpha}}+\Omega_{0B}}}.
\end{eqnarray}
In Fig. \ref{fig10}, with using the age data of the universe, we
plot the isochrones in $\alpha$ - $\omega_{DE}$ plane under the
boundary of domain constrained by $c^{2}_{s}$. Then, we can see that
the parameters $\alpha$ and $\omega_{DE}$ are constrained by the age
of the universe further. Since the luminosity distance for all
models is given by the simple expression for a spatially flat
universe
\begin{equation}
\label{H}
H(z)=\bigg[\frac{d}{dz}\bigg(\frac{D_{L}(z)}{1+z}\bigg)\bigg]^{-1},
\end{equation}
using the Eqs. (\ref{time})-(\ref{H}), we have
\begin{equation}
t_{0}=\frac{D_{L}}{(1+z)^{2}}\bigg|^{\infty}_{0}+\int^{\infty}_{0}\frac{D_{L}}{(1+z)^{3}}dz.
\end{equation}
From above equation, we can see that the luminosity distance is
directly related to the age of the universe.

\subsection{The linear density perturbations}

In this subsection, we will study the growth of density
perturbations for the MGCG fluid in the linear regime on subhorizon
scales. We will follow the analysis method in Ref. \cite{GCG}.
Firstly, we will introduce some quantities used in this part. The
relationship between the comoving coordinate ($\vec{x}$, $t$) and
the physical coordinate ($\vec{r}$, $t$) is $\vec{x}=\vec{r}/a$. The
velocity in physical coordinate is $\vec{u}=\dot{\vec{r}}$.
Furthermore, we can expand $\vec{u}=\dot{a}\vec{x}+\vec{v}$ where
$\vec{v}$ is the first-order perturbation to the Hubble flow
$\dot{a}\vec{x}$.
$\rho_{DM}(\vec{x},t)=\rho_{0DM}(t)(1+\delta_{DM}(\vec{x},t))$ where
$\rho_{0DM}(t)$ is the background density and
$\delta_{DM}(\vec{x},t)$ is the first-order perturbation of
$\rho_{0DM}$. $\phi(\vec{x},t)$ is the first-order perturbation of
Newtonian potential $\Phi$, which satisfies Poisson's equation
$\nabla^2\Phi=4\pi Ga^{2}\rho_{DM}$ in the comoving coordinate.

Before discussing the linear density perturbations, we will research
an explicit interaction between dark matter and dark energy in our
model and depict this interaction through an energy exchange term
$\Gamma$ which will be used in the linear density perturbed
equations. From Eqs. (\ref{rhoDE})-(\ref{rhoDM}), we can obtain the
scaling behavior of the energy densities as follows
\begin{equation}
\frac{\rho_{DM}}{\rho_{DE}}=\frac{\kappa}{\lambda}(1+z)^{-3\omega_{DE}(1+\alpha)}
=\frac{\Omega_{0DM}}{\Omega_{0DE}}(1+z)^{-3\omega_{DE}(1+\alpha)},
\end{equation}
where there is an explicit interaction between dark matter and dark
energy. This can be seen from the energy conservation equation more
clearly, which can be written as
\begin{equation}
\dot{\rho}_{DM}+3H\rho_{DM}=-[\dot{\rho}_{DE}+3H(\rho_{DE}+p_{DE})].
\end{equation}
Furthermore,
\begin{equation}
\dot{\rho}_{DM}+3H\rho_{DM}=\Gamma,
\end{equation}
\begin{equation}
\dot{\rho}_{DE}+3H(\rho_{DE}+p_{DE})=-\Gamma,
\end{equation}
where $\Gamma$ is the source term in the continuity equation due to
the energy transferring between dark matter and quiessence dark
energy in our model. From the above equations, we can respectively
derive the effective equations of state for dark matter
$\omega_{DM}^{eff}$ and dark energy $\omega_{DE}^{eff}$ in the MGCG
scenario
\begin{eqnarray}
\omega_{DM}^{eff}&&=-\frac{\Gamma}{3H\rho_{DM}}\nonumber\\
&&=-\frac{\alpha\omega_{DE}\Omega_{0DE}(1+z)^{3\omega_{DE}(1+\alpha)}}{\Omega_{0DM}
+\Omega_{0DE}(1+z)^{3\omega_{DE}(1+\alpha)}},
\end{eqnarray}
\begin{eqnarray}
\omega_{DE}^{eff}&&=\omega_{DE}+\frac{\Gamma}{3H\rho_{DE}}\nonumber\\
&&=\omega_{DE}(1+\alpha)-\frac{\alpha\omega_{DE}\Omega_{0DE}(1+z)^{3\omega_{DE}(1+\alpha)}}{\Omega_{0DM}
+\Omega_{0DE}(1+z)^{3\omega_{DE}(1+\alpha)}}.
\end{eqnarray}

\begin{figure}[htbp]\center
\includegraphics[height=13cm,angle=270]{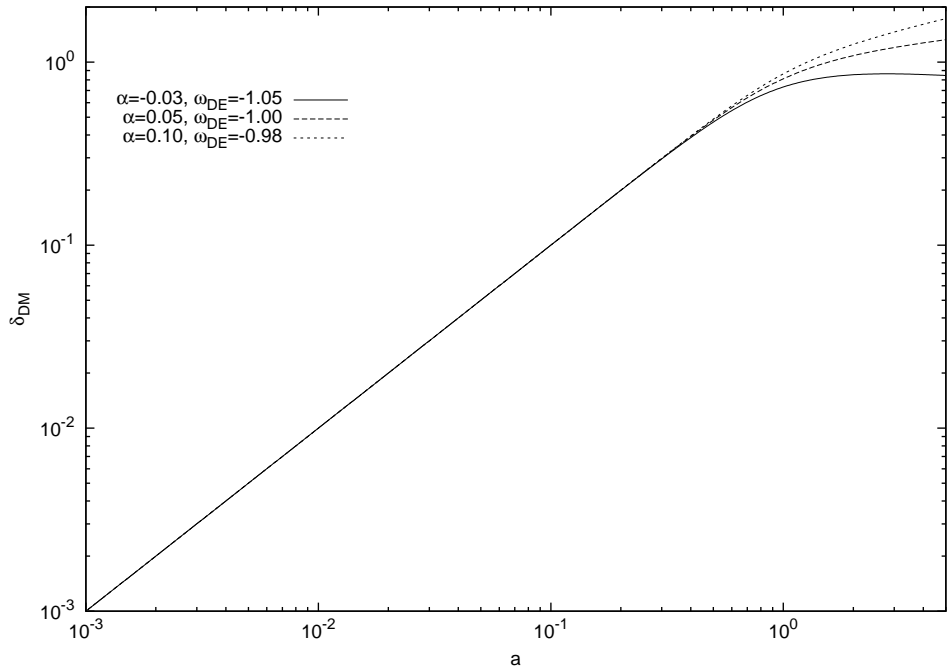}
 \caption{\label{fig11}$\delta_{DM}$ as the function of scale factor
$a$.}
\end{figure}
\begin{figure}[htbp]\center
\includegraphics[height=13cm,angle=270]{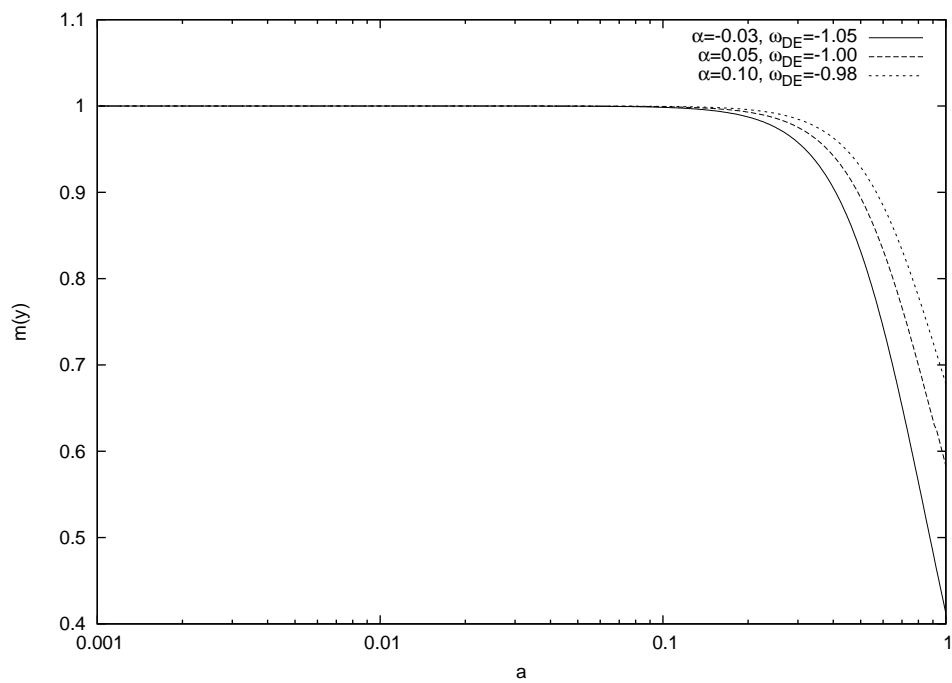}
 \caption{\label{fig12}The growth factor $m(y)$ as a function of
scale factor $a$.}
\end{figure}
\begin{figure}[htbp]\center
\includegraphics[height=13cm,angle=270]{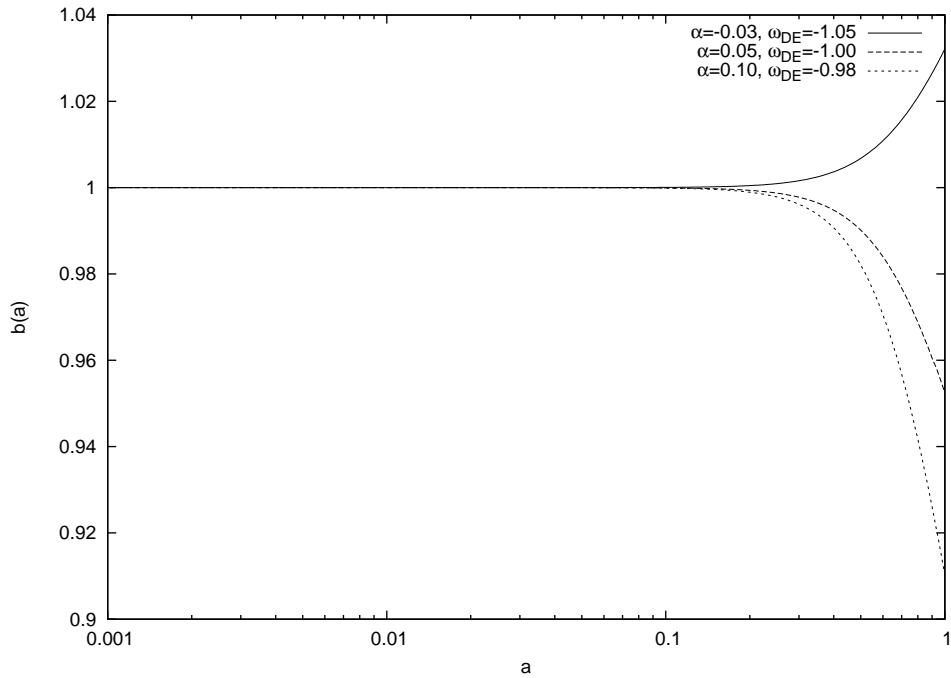}
 \caption{\label{fig13}The bias $b(a)$ as a function of scale factor
$a$.}
\end{figure}
\begin{figure}[htbp]\center
\includegraphics[height=9cm,angle=0]{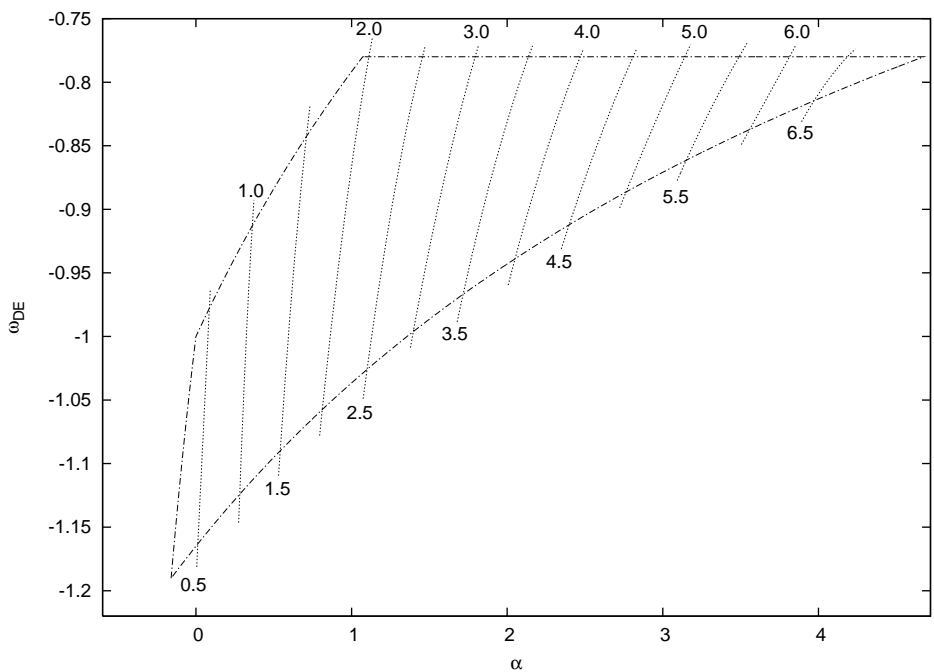}
 \caption{\label{fig14}The contours for growth factor $m$ in the
$\alpha$ - $\omega_{DE}$ plane.}
\end{figure}
\begin{figure}[htbp]\center
\includegraphics[height=9cm,angle=0]{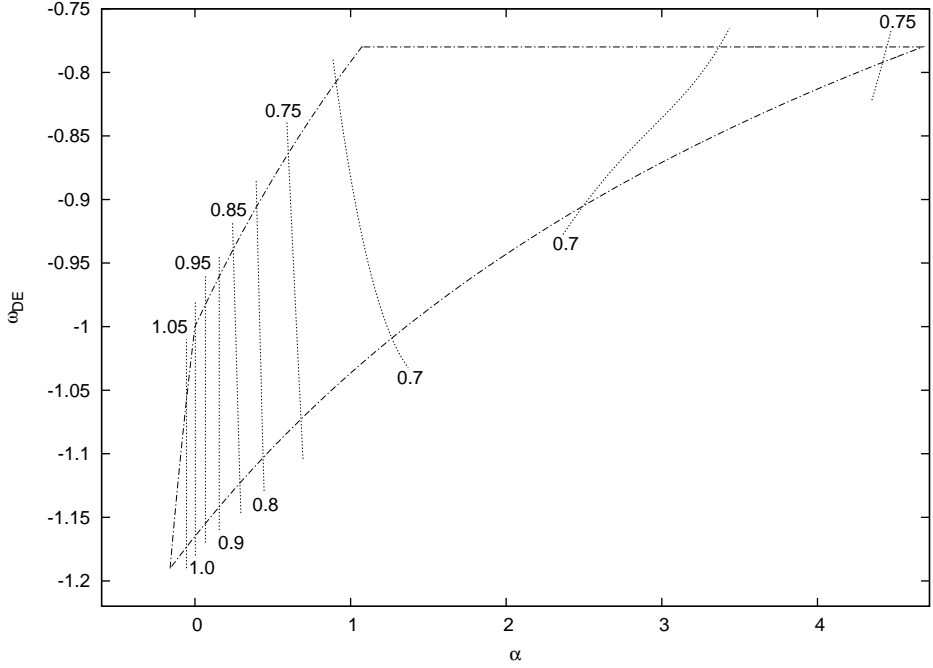}
 \caption{\label{fig15}The contours for bias $b$ in the
$\omega_{DE}$ - $\alpha$ plane.}
\end{figure}
\begin{figure}[htbp]\center
\includegraphics[height=9cm,angle=0]{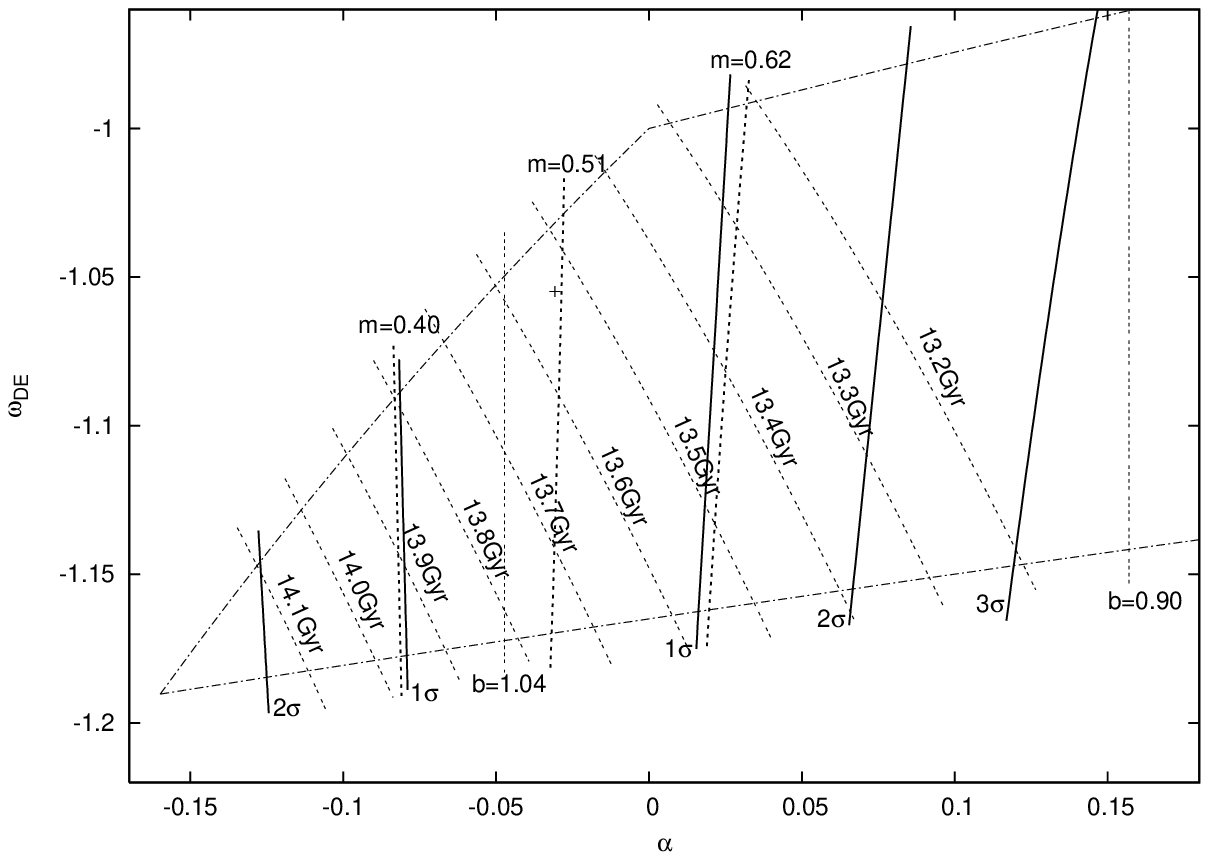}
 \caption{\label{fig16}The contours in solid lines come from four
limits of $c^{2}_{s}$, $t_{0}$, $m$ and $b$ in the $\omega_{DE}$ -
$\alpha$ plane. The confidence level contours of $68.3\%$, $95.4\%$
and $99.73\%$. The $1\sigma$ fit values for our model parameters are
$\alpha=-3.07^{+5.66}_{-4.98}\times10^{-2}$ and
$\omega_{DE}=-1.05^{+0.06}_{-0.11}$. And the plus sign in this
figure denotes the best fits.}
\end{figure}

Now, we focus on the mass density perturbations in the linear
approximation. In this paper, we apply the comoving coordinate
($\vec{x}$, $t$) for a pressureless fluid with background density
$\rho_{0DM}$ and density contrast $\delta_{DM}$, with a source term
$\Gamma$ due to the energy transfer from dark matter to quiessence
dark energy. Based on Ref. \cite{growth}, in the first-order
approximation, the perturbed equations are written in the comoving
coordinate as follows
\begin{equation}
\label{v} \frac{\partial\vec{v}}{\partial
t}+\frac{\dot{a}}{a}\vec{v}=-\frac{\nabla \phi}{a},
\end{equation}
\begin{equation}
\label{delta}
\nabla\cdot\vec{v}=-a\frac{\partial\delta_{DM}}{\partial
t}-a\Gamma\frac{\delta_{DM}}{\rho_{0DM}},
\end{equation}
\begin{equation}
\label{phi} \frac{1}{a^{2}}\nabla^{2}\phi=4\pi
G\rho_{0DM}\delta_{DM},
\end{equation}
where $\delta_{DM}$, $\vec{v}$ and $\phi$ are respectively the
first-order perturbations of $\rho_{0DM}$, $\dot{a}\vec{x}$ and the
Newtonian potential $\Phi$. Since the matter is so cold that the
term proportional to the speed of sound $c^{2}_{s}$ in Eq. (\ref{v})
can be omitted. It leads to disappearance of the term
$\frac{c^{2}_{s}}{a}\nabla\delta_{DM}$ in the Eq. (\ref{v})
\cite{book}. Assuming that both the density contrast $\delta_{DM}$
and peculiar velocity $\vec{v}$ are small, i.e., that
$\delta_{DM}\ll1$ and $\vec{v}\ll \vec{u}$, where $\vec{u}$ is the
velocity in physical coordinate. Taking the divergence of Eq.
(\ref{v}) and substituting Eqs. (\ref{delta}) and (\ref{phi}), we
finally obtain
\begin{eqnarray}
\label{perturbation}
&&H^{2}\delta''_{DM}+\bigg(\dot{H}+2H^{2}+H\frac{\Gamma}{\rho_{DM}}\bigg)\delta'_{DM}-\bigg[4\pi
G\rho_{DM}\nonumber\\
&&-2H\frac{\Gamma}{\rho_{DM}}-H\bigg(\frac{\Gamma}{\rho_{DM}}\bigg)'\bigg]\delta_{DM}=0,
\end{eqnarray}
where $'\equiv d/d\ln a$. Since the term
$\frac{c^{2}_{s}}{a}\nabla\delta_{DM}$ is not included in the Euler
equation, it leads to disappearance of the term
$\nabla^{2}\delta_{DM}$ in Eq. (\ref{perturbation}). And we can
easily see that there is no scale dependent term to drive
oscillations or blowup in the power spectrum. We can solve Eq.
(\ref{perturbation}) numerically, when the initial conditions are
chosen as $a=10^{-3}$ and $\delta_{DM}\simeq10^{-3}$, we plot the
linear density perturbation for dark matter $\delta_{DM}$ as a
function of $a$ (see Fig. \ref{fig11}). In Fig. \ref{fig11}, we can
see the perturbation starts departing from the linear behavior
around $a\simeq0.80$ i.e. $z\simeq0.25$ which is similar to the
epoch when $z<z_{eq}$. Through the energy transferring from dark
matter to dark energy, we can see that the dominance of dark energy
is related to the time when the density fluctuations start deviating
from the linear behavior.

When we consider the period after decoupling, the baryons are no
longer coupled to photons so that the baryons are also a
pressureless fluid like the dark matter. Then, we derive the baryon
perturbation as follows
\begin{equation}
H^{2}\delta''_{B}+(\dot{H}+2H^{2})\delta'_{B}-4\pi
G\rho_{DM}\delta_{DM}=0.
\end{equation}
Then we investigate the growth factor $m(y)$ and the bias $b(a)$
where $m(y)=D'(y)/D(y)$, $D(y)$ is the linear growth function,
$y=\ln(a)$ and $b=\delta_{B}/\delta_{DM}$. One can see from Fig.
\ref{fig12} and Fig. \ref{fig13} that between the present and
$z\simeq5$, the growth factor $m(y)$ and the bias $b(a)$ are quite
sensitive to the values of $\alpha$ and $\omega_{DE}$. Subsequently,
we consider the constraints from $m(y)$ and $b(a)$ on the $\alpha$ -
$\omega_{DE}$ plane. In Fig. \ref{fig14} and Fig. \ref{fig15}, we
have shown the contours for the growth factor $m(y)$ and the bias
$b(a)$ in the $\alpha$ - $\omega_{DE}$ plane under the boundary of
domain constrained by $c^{2}_{s}$. On the other hand, the growth
factor and the bias parameter at $z\simeq0.15$ have been determined
using the 2dFGRS. The redshift space distortion parameter
$\beta=0.49\pm0.09$, and the bias $b=1.04\pm0.11$\cite{mb}. For
$\beta=m/b$, we can subsequently determine the constraint on the
growth factor $m$ as $m=0.51\pm0.11$. Noted that the fitted values
of $b$ and $\beta$ are sensitive to the underlying model and the
observational constraints on them are obtained under $\Lambda$CDM
when converting redshift to distances for the power spectra. The
right way is to conduct data processing for $b$ and $\beta$ by using
2dFGRS data under our model. Then we can constrain $\alpha$ and
$\omega_{DE}$ by using the fitted values. In this paper we use the
above fitted data temporarily and our future work will focus on this
tough work.

In Fig. \ref{fig16}, we consider four restrictions together to give
the best estimation of $\alpha$ and $\omega_{DE}$. Firstly, we plot
contours constraint imposed by $t_{0}$ (Fig. \ref{fig10}), $m$ (Fig.
\ref{fig14}) and $b$ (Fig. \ref{fig15}) under the boundary of domain
constrained by the sound speed. Each set leads to a reduced
$\chi^{2}$: $\chi^{2}(\alpha,\omega_{DE})=$(Theory
values-Observation values)$^{2}$/(Observational error)$^{2}$. To
combine the constraints on $\alpha$ and $\omega_{DE}$ coming from
the age of the universe $t_{0}$, growth factor $m$ and the bias $b$
within the domain given by the limits of the sound speed
$c^{2}_{s}$, we have added their individual $\chi^{2}$ as if they
were part of a total experiment with uncorrelated Gaussian errors:
$\chi^{2}_{total}(\alpha,\omega_{DE})=\chi^{2}_{t_{0}}(\alpha,\omega_{DE})
+\chi^{2}_{m}(\alpha,\omega_{DE})+\chi^{2}_{b}(\alpha,\omega_{DE})$.
Therefore, we plot the contour level
$\Delta\chi^{2}_{total}(\alpha,\omega_{DE})=2.3$,
$\Delta\chi^{2}_{total}(\alpha,\omega_{DE})=6.17$ and
$\Delta\chi^{2}_{total}(\alpha,\omega_{DE})=11.8$, where
$\Delta\chi^{2}_{total}(\alpha,\omega_{DE})=\chi^{2}(\alpha,\omega_{DE})-(\chi^{2}(\alpha,\omega_{DE}))_{min}$,
defines, respectively, for two degrees of freedom the $68.3\%$,
$95.4\%$ and $99.73\%$ confidence levels represented in Fig.
\ref{fig16}. The $1\sigma$ fit values are
$\alpha=-3.07^{+5.66}_{-4.98}\times10^{-2}$ and
$\omega_{DE}=-1.05^{+0.06}_{-0.11}$. Noted that the presence of the
cutoff contours' curves in Fig. \ref{fig16} is due to the restricted
area boundary which comes from the limits of $c^{2}_{s}$.

\section{Conclusion}
A new model named as the modified generalized Chaplygin gas (MGCG),
which is a further generalization of the generalized Chaplygin gas,
has been proposed in the present paper. As a version of the unified
dark matter and dark energy, this MGCG fluid is consisted of dark
matter and quiessence dark energy with constant $\omega_{DE}$.
Firstly, fundamental cosmology equations for the MGCG have been
described. For tests with future deep observations, we then consider
the Statefinder diagnostic since it can probe the expansion dynamics
of the universe through higher derivatives of the scale factor.
Furthermore, we investigate the evolution of density perturbations
and the structure formation in our model. Then, with applied the age
of the universe data $t_{0}$, growth factor $m$ and the bias $b$
within the domain given by the limits of $c^{2}_{s}$, the parameters
$\alpha$ and $\omega_{DE}$ are constrained:
$\alpha=-3.07^{+5.66}_{-4.98}\times10^{-2}$ and
$\omega_{DE}=-1.05^{+0.06}_{-0.11}$. It has been shown that there
are a little difference between our model and others. To determine
whether our model could be a candidate model of dark energy, the
analysis and discussion of the matter power spectrum for MGCG must
be done. This is a subject of our future research.

\section*{Acknowledgments}
 X.-M. Deng thanks the suggestions and comments from a referee,
which plays an important role for the improvement of this article.
X.-M. Deng appreciates the support from the group of Almanac and
Astronomical Reference Systems in the Purple Mountain Observatory of
China.

\end{document}